\documentclass[11pt]{article}

\DeclareMathAlphabet{\mathpzc}{OT1}{pzc}{m}{it}

\usepackage{latexsym}
\usepackage{marvosym}
\usepackage{amssymb}
\usepackage{amsfonts}
\usepackage{amsmath}
\usepackage{amsthm}
\usepackage{dsfont}    
\usepackage{bbold}     
\usepackage[english]{babel}
\usepackage{caption}
\usepackage{epsfig}
\usepackage{float}
\usepackage{subfigure}
\usepackage{psfrag}
\usepackage{graphicx}
\usepackage{epsfig}
\usepackage{relsize}
\usepackage{comment}
\usepackage{enumerate}

\newcommand{\dd}{\mathrm{d}}
\newcommand{\ee}{\mathrm{e}}
\newcommand{\ii}{\mathrm{i}}

\newcommand{\ceff}{C_{\mathrm{eff}}}

\newcommand{\mep}{\langle p\rangle}
\newcommand{\meps}{\langle p\rangle_s}
\newcommand{\mepl}{\langle p\rangle_l}
\newcommand{\mn}[1]{\langle #1\rangle}

\newcommand{\mepkw}{\langle p(\omega)\rangle_{pk}}
\newcommand{\rpk}{R_{pk}}
\newcommand{\re}{\operatorname{Re}}

\newtheorem{property}{Property}

\newcommand{\e}{\varepsilon}
\newcommand{\al}{\alpha}

\begin{document}

\title{Power computation for the triboelectric nanogenerator}
\author{J.H.B.\ Deane$^1$, R.D.I.G.\ Dharmasena$^2$ and G. Gentile$^3$\\
$^1$Department of Mathematics, University of Surrey, Guildford, GU2 7XH, UK\\
$^2$Advanced Technology Institute, University of Surrey, Guildford, GU2 7XH, UK\\
$^3$Dipartimento di Matematica, Universit\`{a} Roma Tre, Roma, I-00146, Italy}


\date{}
\maketitle

\begin{abstract}
We consider, from a mathematical perspective, the power generated by a contact-mode
triboelectric nanogenerator, an energy harvesting device that has been thoroughly
studied recently. We encapsulate the behaviour of the device in a differential equation, which
although linear and of first order, has periodic coefficients, leading
to some interesting mathematical problems. In studying these, we derive
approximate forms for the mean power generated and the current waveforms, and
describe a procedure for computing the Fourier coefficients for the current,
enabling us to compute the power accurately and show how the power is distributed
over the harmonics.  Comparisons with numerics validate our analysis.
\end{abstract}

\section{Introduction}
Triboelectric nanogenerators (TENGs) have received
considerable attention recently as potential candidates for energy
scavenging~\cite{envsci17, id1, id2, id4}. These devices have been
shown to convert mechanical energy into electricity in
applications such as energy harvesting and self-powered sensors~\cite{id5,
id6, id7}. Furthermore, TENGs have many advantages over existing energy harvesting
technologies~\cite{id1, id2, id5, id7, id8}, such as low cost, simple
construction, relatively high power, flexibility and robustness.

The contact-mode triboelectric nanogenerator is the most
commonly used TENG architecture owing to its simplicity and output
performance~\cite{id8, id9}. Typically, it consists of two triboelectric
plates, at least one being of dielectric material, each attached to an
electrode. When the plates come into contact, one becomes
positively charged and the other, negatively. (Static electricity produced
by friction is a well-known example of the same effect.) We feel that it is timely
to discuss, from an applied mathematical point of view, the power produced by a TENG,
whose construction is described in detail in for example~\cite{envsci17, id4, id8}.

A related device, a piezo-electric generator designed to harvest energy
from the heartbeat, is described in~\cite{zhang}, but its mathematical
modelling, at least from the point of view we take here, is straightforward and
so of less interest.

\begin{figure}[htbp]
\begin{centering}
\includegraphics*[width=4in]{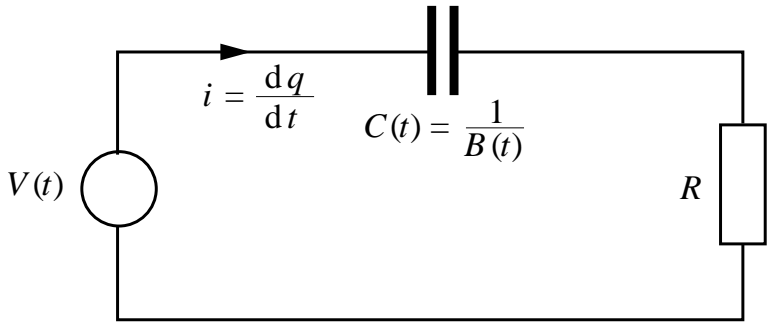}
\caption{A circuit model of the TENG studied in this paper, in which the voltage
source $V(t)$ and the capacitance $C(t)$ are both periodic with the same
fundamental frequency $\omega$. This circuit is described by the
differential equation~\eqref{ode}.}
\label{Matt}
\end{centering}
\end{figure}

In this paper, we consider the most common configuration --- two metal electrodes, each with
a layer of dielectric attached --- in order to assess its power output characteristics.
Our starting point is the ordinary differential equation (o.d.e.) that describes such a TENG
connected to a load resistance $R$. Our main assumption is that the TENG is
being driven periodically at a frequency $\omega$, that is to say, the
separation of the plates varies periodically with time. We then adopt a circuit theory approach, as
laid out in~\cite{wang_ode}, by modelling the system as the circuit in Figure~\ref{Matt}.
The circuit leads directly to a differential equation, and this forms the
basis for our study. The circuit and its mathematical description are both
straightforward, the only complication arising from the fact that the
capacitance, $C(t)$, is a periodic function of time: this is inescapable and is a direct
consequence of the periodically varying plate separation. It is this that generates the
time variation in both the capacitance and the input voltage, and
guarantees that their fundamental frequencies, $\omega$, are identical.

We are particularly interested in computing, both accurately and approximately, the mean
power delivered to the load as a function of the parameters of the system, and in order to
investigate this, we first take a perturbation theory approach. This leads to
two formulae for the power, one valid for small $R$ and the other for large
$R$. We also tackle the same problem via Fourier series, this approach showing
how the power generated is distributed over different harmonics, that is,
integer multiples of $\omega$. The mean (as opposed to, say, the peak) power
is especially amenable to calculation, and is also the most appropriate measure of the
effectiveness of the generator.

We should give here some justification for our analytical approach, which
takes up the bulk of the paper. After all, the mean power, current
waveforms and so on can also be computed numerically --- why, then, compute
them analytically? We offer several reasons for our approach. In general,
analytical results give deeper insight into the problem, and in this paper,
we derive several approximate expressions for the variables of interest,
and use these, for instance, to optimise the power output by means of simple arguments.
For example, we derive, by perturbation theory, a simple expression for the value 
of $R$ that maximises the mean power. Contrast this with a purely numerical approach, 
in which such an optimisation would have to be carried out for one set of parameters at a time. 
Furthermore, for large $R$, the transient times are large --- a
numerical solution would have to be continued for long times in order to
ensure that the transient has decayed sufficiently. By contrast, our
analytical solution is set up specifically to correspond to the steady state (post-transient)
behaviour. The Fourier series approach, which we also discuss, immediately gives
insight into how rapidly the Fourier coefficients decrease in magnitude
with index. It also turns out to be possible to express these coefficients explicitly in terms of
Bessel functions.

The rest of the paper is organised as follows.  We first derive the o.d.e., including the 
periodic functions $V(t)$ and the reciprocal capacitance (elastance), $B(t) := 1/C(t)$.
Both of these depend on a function $I(x)$, which in turn comes from calculating the
electric field in the system. This was derived in~\cite{envsci17} for
square plates; for the sake of completeness, we give the derivation in the
slightly more general case of rectangular plates in Appendix~I.
On examining both $V(t)$ and $B(t)$ for practical values of the parameters,
we make simple approximations to them, which in turn leads to a simplified
version of the o.d.e. We then carry out the mean power calculations mentioned above,
in both the case of general periodic source voltage and elastance, and their
approximations, and follow this with some comparisons between numerical solutions
of the o.d.e.\ in both cases. This comparison shows the approximations to be
very good. We then discuss an approach to power computation
based on Fourier series, after which we draw some conclusions.

\section{The o.d.e.}

The o.d.e.\ in its original form
\begin{equation}
AR\dot{\sigma} + \underbrace{\frac{1}{\pi}\left(\frac{1}{\epsilon_1} +
\frac{1}{\epsilon_2}\right)\left\{I(x_1 + x_2 + z(t)) - I_0\right\}}_{G(t)}\,\sigma = 
F_1(t) + F_2(t)
\label{ode_orig}
\end{equation}
was discussed in~\cite{envsci17}. Here,
$\sigma(t)$, to be solved for, is the surface charge density and $\dot\sigma$ is its time
derivative. Also, $A = wl = \alpha w^2$ is the area of the plates, $\alpha = l/w$ is their aspect ratio,
$R > 0$ is a fixed load resistance, $\epsilon_1$ and $\epsilon_2$ are the permittivities
of the dielectrics, with thickness $x_1$ and $x_2$ respectively;
and $\sigma_T$ is the triboelectric charge density, which is constant and a
property of the materials used. Furthermore, the excitation $z(t) =
z_0[1+\sin(\omega t +\phi)]$, where $z_0$ is the amplitude, $\phi$ is a fixed phase angle, and
$\omega = 2\pi f$ is the excitation frequency; and
$F_i = \frac{\sigma_T}{\pi\epsilon_i}\left\{I\left(x_i + z(t)\right) - I(x_i)\right\}$,
with $i = 1,2$, where $I(x)$ is defined in equation~(\ref{Idef}) in 
Appendix~I. Finally, $I_0 = \lim_{x\rightarrow 0} I(x)$. The variable parameters of 
interest are $R$ and $\omega$ --- we consider all the other parameters to be fixed.

Since $\sigma$ is a charge density, we interpret $q := A\sigma$ as a charge.
Hence, replacing $A\sigma$ in equation~(\ref{ode_orig}) with $q$, we have
$$
R\dot{q} + \frac{G(t)}{A} q = F_1(t) + F_2(t),
$$
where $G(t)$ is defined in equation~(\ref{ode_orig});
and it becomes clear that the function of time that multiplies $q$ can be
interpreted as the reciprocal of a time-dependent capacitance, provided that
this is defined as the ratio of the change in the charge in the system to the change
in potential difference (as opposed to the derivative of the charge with respect to potential
difference); and that the term on the right hand
side is a time-dependent voltage, $V(t) = F_1(t) + F_2(t)$. That is,
\begin{equation}
R\dot{q} + B(t) q = V(t)
\label{ode}
\end{equation}
where
\begin{equation}
B(t) = \frac{G(t)}{A} = \frac{1}{A\pi}\left(\frac{1}{\epsilon_1} + \frac{1}{\epsilon_2}\right)
\left\{I(x_1 + x_2 + z(t)) - I_0\right\}.
\label{Bdef}
\end{equation}
We claim that equation~(\ref{ode}) models the TENG and the rest of this paper
discusses its periodic solution, $q(t)$, from which the current, $i(t) := \dot{q}(t)$,
and much else, can be deduced.

\subsection{Practical parameter values}

\begin{figure}[htbp]
\centering 
\includegraphics*[width=5in]{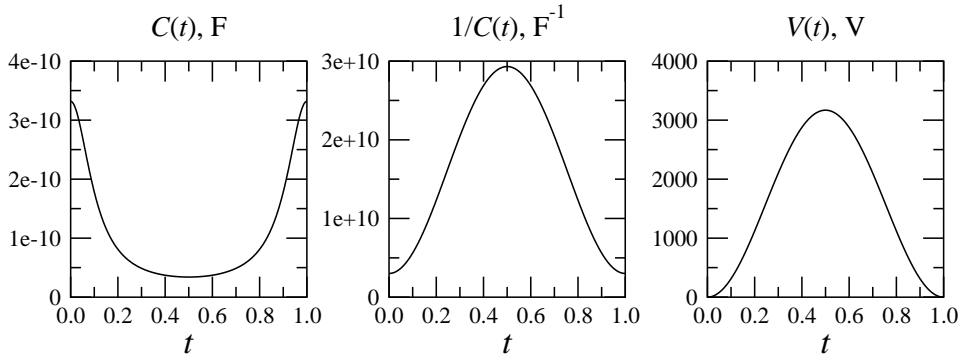}
\caption{The periodic functions of time in equation~(\ref{ode}), for
$\omega = 2\pi$ rad/s: $C(t)$, left; $B(t) = 1/C(t)$, middle; and $V(t)$, right.
None of the functions are approximated. Parameter values are from Table~\ref{numvals}.}
\label{CV}
\end{figure}

Typical values for the parameters, taken from~\cite{envsci17}, are given in
Table~\ref{numvals}. Plots of the exact $C(t)$, $B(t)$ and $V(t)$ are given in Figure~\ref{CV} for these
parameter values. Note that, despite the fact that $z(t)$ contains only one
harmonic, $C(t)$ and $V(t)$ contain all harmonics owing to the nonlinear
function $I(\mbox{const.}+z(t))$ used in their definition. However, Figure~\ref{CV}
suggests that a good approximation might be obtained by only considering
the first harmonic.

\begin{table}[ht]
\centering
\begin{tabular}{lll} \hline
\multicolumn{3}{c} {\raisebox{2.2ex}{ }Parameter values for practical triboelectric
nanogenerator}\\ \hline
Name & Symbol & Numerical value\\ \hline
\raisebox{2.2ex}{}Permittivity of free space    & $\epsilon_0$ & $8.854\times 10^{-12}$ Fm$^{-1}$\\ \hline
Permittivity of dielectric 1   & $\epsilon_1$ & $3.30\epsilon_0$\\
Permittivity of dielectric 2   & $\epsilon_2$ & $3.27\epsilon_0$\\
Triboelectric charge density   & $\sigma_T$ & $4.8\times 10^{-5}$ Cm$^{-2}$\\
Thickness of dielectric 1   & $x_1$ & 200 $\mu$m\\
Thickness of dielectric 2   & $x_2$ & 20 $\mu$m\\ \hline
Default aspect ratio, $l/w$ & $\alpha$ & 1\\
Default plate dimensions   & $l, w$  & $l = w = 5\times 10^{-2}$ m\\
Default plate area & $A$ & $2.5\times 10^{-3}$ m$^2$\\
Excitation amplitude \& phase & $z_0,\,\phi$ & resp.\  $1.0\times 10^{-3}$ m, $3\pi/2$\\
Elastance parameters & $B_0, B_1$ & resp.\ $1.6\times 10^{10}, 1.3\times 10^{10}$ F$^{-1}$\\
Drive voltage amplitude & $V_0$ & 1550 V\\ \hline
Excitation frequency & $f = \omega/2\pi$ & 0.1 -- $10^3$ Hz (nom.\ 1 Hz)\\
Load resistance & $R$ & $10^5$ -- $10^{15}\Omega$\\ \hline
\end{tabular}
\caption{Names, symbols and numerical values for the parameters for a practical TENG.
Note the very large range of the parameter $R$.}
\label{numvals}
\end{table}

\subsection{The practical approximation}

Clearly, $I(\mathrm{constant} + z(t))$ is not sinusoidal, even
though $z(t)$ is: so strictly speaking, $B(t)$ and $V(t)$ should be
expanded as Fourier series, both with identical fundamental frequency $\omega$.
However, in the practical case considered here, $x_1$, $x_2$, $z_0$ and $I$ are such that good first
approximations for $B(t)$ and $V(t)$ are
\begin{equation}
B(t) = \sum_{k\in\mathbb Z}b_k \ee^{\ii k\omega t} \approx B_0 - B_1 \cos\omega t\qquad\qquad
V(t) = \sum_{k\in\mathbb Z}v_k \ee^{\ii k\omega t}\approx V_0(1-\cos\omega t),
\end{equation}
as suggested by Figure~\ref{CV}. In fact,
$\left|b_2/b_1\right| \approx 9.385\times 10^{-3},$
$\left|b_3/b_1\right| \approx 7.610\times 10^{-6}$ and
$\left|v_2/v_1\right| \approx 9.351\times 10^{-3}$,
$\left|v_3/v_1\right| \approx 6.898\times 10^{-6}.$
Thus, we study where appropriate both
the general o.d.e.\ with $B(t)$ and $V(t)$ subject to some mild conditions but
otherwise arbitrary, and also the approximate o.d.e.\
\begin{equation}
R\dot{q} + (B_0 - B_1\cos\omega t)q = V_0(1 - \cos\omega t).
\label{apode}
\end{equation}
We refer to equation~\eqref{apode} as the `practical approximation' in what follows.
A better approximation would take into account more terms in the Fourier
series expansions of $B(t)$ and $V(t)$, but in practice, for the parameter
values given in Table~\ref{numvals}, this approximation gives good results ---  as we shall see.

\section{Analytical results}

\subsection{Unique periodic attractor}
\label{exist}

In this section, we show that, under conditions that must apply on physical
grounds, there is a unique periodic solution to the differential
equation~\eqref{ode}, and that solutions starting from any initial
condition are attracted to it. We can then talk about `the periodic
solution', knowing that this exists.

In a real system described by the differential equation~\eqref{ode},
on physical grounds alone, $B(t)$ and $V(t)$ must satisfy the Dirichlet
conditions~\cite{haykin},
and so both functions can be expanded in Fourier series. Furthermore, as
shown in the Appendix~I, $B(t) > 0$ for all $t$ and so the mean value of $B(t) > 0$.
We will use both these facts in what follows.

In practice, we will be interested only in periodic solutions to the
o.d.e.~\eqref{ode}, but in this section alone, we need to solve the o.d.e.\ for the complete
solution, which we call $q_c(t)$. The standard way to solve such an o.d.e.\ is by the
integrating factor method~\cite{piaggio}, which gives 
\begin{equation}
q_c(t) = \exp\left\{-\frac{1}{R}\int_0^t B(t')\,\dd t'\right\}
\left[\frac{1}{R}\int_0^t V(t') \exp\left\{\frac{1}{R}\int_0^{t'} B(t'')\,\dd t''\right\}
\dd t' + q_c(0)\right].
\label{intfac}
\end{equation}
Although the solution in this form is not obviously useful for direct
computation of, for instance, the mean power, it \textit{is} useful to show 
that for any initial condition, $q_c(0)$, there is a unique periodic attractor. We now prove this.

We introduce the notation for the mean, $\mn{f}$, of any periodic function $f(t)$
with period $T_0$, which is
$$\langle f\rangle := \frac{1}{T_0}\int_{-T0/2}^{T_0/2} f(t)\,\dd t.$$
We then define $\widetilde{f}(t) := f(t) - \mn{f}$ as the zero-mean, time-varying part of
$f(t)$.

Since $B(t)$ and $V(t)$ both have the same period $T_0 = 2\pi/\omega$, all
the time-varying terms in equation~\eqref{intfac} have period $T_0$. Hence,
writing $B(t) = \mn{B} + \widetilde{B}(t)$,
we can make the following Fourier expansion:
$$\frac{V(t)}{R} \exp\left\{\frac{1}{R}\int_0^{t} B(t')\,\dd t'\right\} =
\ee^{\mn{B}t/R}\,\frac{V(t)}{R}\exp\left\{\frac{1}{R}\int_0^{t} \widetilde{B}(t')\,\dd t'\right\} =
\ee^{t/\tau}\sum_{k\in\mathbb Z} I_k \ee^{\ii k\omega t},$$
where $\tau = R/\mn{B} > 0$ since both $R$ and $\mn{B}$ are positive, and Fourier 
coefficients $I_k\in\mathbb C$ have dimensions of current.  Using this in~\eqref{intfac} then gives
\begin{eqnarray*}
q_c(t) &=& \ee^{-t/\tau}\,\exp\left\{-\frac{1}{R}\int_0^t \widetilde{B}(t')\,\dd t'\right\}
\left[\int_0^t\sum_{k\in\mathbb Z} I_k \ee^{(1+\ii k\omega\tau)t'/\tau}\,\dd t'+ q_c(0)\right]\\
&=& \exp\left\{-\frac{1}{R}\int_0^t \widetilde{B}(t')\,\dd t'\right\}
\left[\sum_{k\in\mathbb Z}\frac{\tau I_k\ee^{\ii k\omega t}}{1+\ii k\omega\tau}
+ \ee^{-t/\tau}\left(q_c(0) - \sum_{k\in\mathbb Z}\frac{\tau I_k}{1+\ii k\omega\tau}\right)\right].
\end{eqnarray*}
From this we deduce that
\begin{enumerate}
\item Since $\tau > 0$, as $t\rightarrow\infty$, for all initial conditions $q_c(0)$,
$q_c(t)$ tends to the period-$T_0$ function
$$q(t) = \exp\left\{-\frac{1}{R}\int_0^t \widetilde{B}(t')\,\dd t'\right\}
\sum_{k\in\mathbb Z}\frac{\tau I_k\ee^{\ii k\omega t}}{1+\ii k\omega\tau};$$
\item Any solution $q_c(t)$ approaches this periodic attractor 
exponentially,\footnote{Although the rate of approach to the attractor will be very
slow for large $R$ (i.e. large $\tau$).}
that is, $|q_c(t) - q(t)| < c \ee^{-t/\tau}$ for some positive constant $c$.
\item The choice $q_c(0) = \sum_{k\in\mathbb Z}\tau I_k/(1+\ii k\omega\tau)$
puts the solution directly on the periodic attractor.
\end{enumerate}

\subsection{Perturbation series --- small $R$}

Although the previous section derives an exact expression for the complete
solution, $q_c(t)$, and the periodic solution, $q(t)$, these
expressions do not lend themselves to direct computation of the power, or
simple approximations to it. To circumvent
this, we estimate $q(t)$ using a perturbation theory approach.
Underlying this approach is the assumption that $B(t)$ and $V(t)$ are
both infinitely differentiable (which is clearly the case for the practical approximation).

Our starting point is to define the dimensionless parameter $\e :=
\omega R/\mn{B}$, which is small for small $R$. In terms of this, we
re-write the o.d.e.~\eqref{ode} as
\begin{equation}
\e\dot{q} + \frac{\omega}{\mn{B}}B(t)\, q = \frac{\omega}{\mn{B}} V(t).
\label{Rsode}
\end{equation}
We now set
\begin{equation}
q(t) = q_0(t) + \e q_1(t) + \e^2 q_2(t) + \ldots,
\label{pertser}
\end{equation}
and substituting this into~\eqref{Rsode}, and equating to zero the
coefficients of each power of $\e$, we find
$$\e^0:\;\; q_0 = \frac{V}{B},\;\;\; \e^1:\;\;\dot{q}_0 + \frac{\omega}{\mn{B}}\, B q_1 = 0,
\;\;\; \e^2:\;\;\dot{q}_1 + \frac{\omega}{\mn{B}}\, B q_2 = 0,\ldots,\;
\e^k:\;\;\dot{q}_{k-1} + \frac{\omega}{\mn{B}}B q_k = 0,$$
where for brevity we have dropped the argument $(t)$ for $B$, $V$ and $q_k$.
Hence, we have
$$q_0 = \frac{V}{B},\qquad 
q_1 = -\frac{\mn{B}}{\omega B}\,\frac{\dd}{\dd t}\!\left(\frac{V}{B}\right)
= -\frac{\mn{B}}{\omega}\;\frac{B\dot{V} - \dot{B}V}{B^3},$$
$$q_2 = -\frac{\mn{B}}{\omega B}\,\dot{q}_1 = 
\frac{\mn{B}^2}{\omega^2}\frac{1}{B}\;\frac{\dd}{\dd t}\left(\frac{1}{B}\frac{\dd}{\dd
t}\!\left(\frac{V}{B}\right)\right) =
\frac{\mn{B}^2}{\omega^2}\;\frac{B(B\ddot{V}-\ddot{B}V) -
3\dot{B}(B\dot{V}-\dot{B}V)}{B^5} $$
with obvious generalisations for $q_k$, $k > 2$.  The above expressions for $q_k$ are 
general, in that they apply for any infinitely differentiable functions $B, V$ provided 
only that $B > 0,\,\forall t$.

The power series expansion~\eqref{pertser} is very unlikely to converge. Even in
the case of constant $B$, it is easy to check that the periodic solution $q(t)$ is not analytic
around $\e = 0$ for a generic, analytic $V(t)$ containing all the harmonics. Additionally, the
presence of the function $B(t)$ will only complicate the situation. However, even
though the series expansion is not expected to be convergent, it is likely to be an asymptotic series,
and hence any truncation of it should provide a reliable approximation of the full solution, for
$\e$ small enough.

We can make further progress from this point if we use the practical approximation,
which is that $B(t) = B_0 - B_1\cos\omega t$ and $V(t) = V_0(1 - \cos\omega t)$,
where $B_0 = \mn{B} > B_1 > 0$. With these definitions of $B$ and $V$ in
force, we have that, for $k\in\mathbb N$, $q_{2k-1}$ are odd and $q_{2k}$ are
even functions of $t$; hence, $\dot{q}_{2k-1}$ are even and $\dot{q}_{2k}$ are odd.
The instantaneous power $p(t) := \dot{q}(t)^2R$ and the mean power is defined as $\mn{p} := \mn{\dot{q}^2}R$. From the
perturbation series~\eqref{pertser}, we have
$$\mn{\dot{q}^2} = \mn{\dot{q}_0^2} + 2\e\mn{\dot{q}_0\dot{q}_1} +
\e^2(2\mn{\dot{q}_0\dot{q}_2} + \mn{\dot{q}_1^2}) + O(\e^3).$$
Now, by the definition of $\mn{f}$ we have that $\mn{f} = 0$ if $f$ is an odd function of $t$. Thus,
using the parity of functions $q_k$ and their first derivatives, we have that
$\mn{\dot{q}_0\dot{q}_1} = 0$ and so on. Hence, for small $R$, and so for small $\e$, we have
$$\mn{p}_s = R\mn{\dot{q}_0^2} + R\e^2 \left(2\mn{\dot{q}_0\dot{q}_2} + \mn{\dot{q}_1^2}\right)
+ O\left(\e^4\right).$$

Since it is useful to know the dependence of $\mn{p}$ on $\omega$ as well
as $R$, we compute
$$\mn{\dot{q}_0^2} = \frac{1}{T_0}\int_{-T_0/2}^{T_0/2}\frac{(B\dot{V} - \dot{B}V)^2}{B^4}\,\dd t = 
\frac{\omega^2}{2\pi}\int_{-\pi}^{\pi} \frac{(B V' - B' V)^2}{B^4}\,\dd x
:= \omega^2 a_1,$$
where we have substituted $x = \omega t$ and used a prime to denote
differentiation with respect to $x$.
In fact, this integral can be evaluated in closed form; it is
\begin{equation}
a_1 = \frac{V_0^2 B_0}{2\sqrt{B_0^2 - B_1^2}(B_0 + B_1)^2}.
\label{a1xp}
\end{equation}
From our previous definitions of $q_0$, $q_1$ and $q_2$, we also have
$$\frac{2}{\omega^2 B_0^2}\mn{\dot{q}_0\dot{q}_2} = \frac{1}{\pi}\int_{-\pi}^\pi
\left(\frac{V}{B}\right)'\frac{\dd}{\dd x}\left\{\frac{1}{B}\frac{\dd}{\dd x}\left[\frac{1}{B}
\left(\frac{V}{B}\right)'\right]\right\}\,\dd  x$$
and
$$\frac{1}{\omega^2 B_0^2}\mn{\dot{q}_1^2} = \frac{1}{2\pi}
\int_{-\pi}^\pi\left\{\frac{\dd}{\dd x}\left[\frac{1}{B}\left(\frac{V}{B}\right)'\right]
\right\}^2\,\dd x.$$
Defining $a_3$ as the sum of the right hand sides of these two expressions,
we can also determine that
$$a_3 = \frac{V_0^2 B_0\left(8 B_0^4 + 44 B_0^2B_1^2 + 17 B_1^4\right)}
{16\sqrt{B_0^2 - B_1^2}(B_0-B_1)^3(B_0+B_1)^5}.$$

Bearing in mind now that $\e = \omega R/B_0$, and that $a_1, a_3$ do not depend on
$\omega$ or $R$, we have that
\begin{eqnarray}
\mn{p}_s &=& a_1\omega^2 R + a_3\omega^4 R^3 + O\left(\omega^6 R^5\right)\nonumber \\
&=& 2.45018\times 10^{-15} \omega^2 R - 1.35505\times 10^{-33}\omega^4 R^3 + O\left(\omega^6
R^5\right)
\label{psmall}
\end{eqnarray}
where $a_1$ and $a_3$ have been evaluated using the values in Table~\ref{numvals}.

\subsection{Perturbation series --- large $R$}

In the case of large $R$, we compute a solution to the o.d.e.\ in a similar way
to the previous section, this time expanding as a power series
in the dimensionless, small parameter $\delta := \mn{B}/\omega R\; (= 1/\e)$. In
terms of $\delta$, the o.d.e.~\eqref{ode} becomes
\begin{equation}
\dot{q} = \delta\frac{\omega}{\mn{B}}\left[V(t) - q\,B(t)\right].
\label{Rlode}
\end{equation}
Again, we expand $q(t) = q_0(t) + \delta q_1(t) + \delta^2 q_2(t) + \delta^3 q_3(t) +\ldots$
and, substituting this into~\eqref{Rlode} and matching powers of $\delta$, we find
$$\delta^0:\;\; \dot{q}_0 = 0,\qquad 
\delta^1:\;\;\dot{q}_1 = \frac{\omega}{\mn{B}}\left(V - q_0\,B\right),\qquad
\delta^2:\;\;\dot{q}_2 = -\frac{\omega}{\mn{B}}\, B\, q_1,\qquad
\delta^3:\;\;\dot{q}_3 = -\frac{\omega}{\mn{B}}\, B\, q_2$$
and so on.

The guiding principle for finding $q_k$ is that for each $k$, $q_k$ is periodic.
Hence, when $q_k$ is expressed as an integral, we must ensure that the
integrand has, in each case, zero mean (otherwise the integral will grow
linearly with $t$).

The equation $\dot{q}_0 = 0$ has solution $q_0 = c_0$ for $c_0$ an as yet
undetermined constant. Hence, we have $\dot{q}_1 = \omega(V - c_0 B)/\mn{B}$.
Recalling that $B(t) = \mn{B} + \widetilde{B}(t)$ and $V(t) = \mn{V} + \widetilde{V}(t)$,
we find
$$q_1(t) = \frac{\omega}{\mn{B}}\int_0^t\mn{V}-c_0\mn{B} +\widetilde{V}(t')
-c_0\widetilde{B}(t')\;\dd t' + c_1,$$
where $c_1$ is a constant to be determined at the next stage. Furthermore,
in order for the integrand here to have zero mean, we require $c_0 = \mn{V}/\mn{B}$. Hence,
$$q_1(t) = \frac{\omega}{\mn{B}^2}\int_0^t \mn{B}\widetilde{V}(t') - \mn{V}\widetilde{B}(t')\,\dd t'+c_1
\qquad\mbox{and}\qquad
\dot{q}_1(t) = \frac{\omega}{\mn{B}^2}\left(\mn{B}\widetilde{V}(t) - \mn{V}\widetilde{B}(t)\right).$$
The constant $c_1$ is found in an analogous way to $c_0$. Since
$\dot{q}_2(t) = -\omega B(t) q_1(t)/\mn{B}$, we have, using our expression for $q_1(t)$,
\begin{equation}
q_2(t) = -\frac{\omega}{\mn{B}}\int_0^t \left(\mn{B}+\widetilde{B}(t')\right)
\left\{\int_0^{t'}\frac{\omega}{\mn{B}^2}\left(\mn{B}\widetilde{V}(t'')
- \mn{V}\widetilde{B}(t'')\right)\,\dd t'' +c_1\right\}\,\dd t' + c_2,
\label{q2def}
\end{equation}
where $c_2$ is a new constant that can be determined at the next stage.
Reasoning as before, for $q_2(t)$ to be periodic, the integrand of the
integral w.r.t.\ $t'$ in
equation~\eqref{q2def} must have mean zero, which fixes $c_1$ by
$$c_1 = -\frac{\omega^2}{2\pi\mn{B}^3}\int_{-T_0/2}^{T_0/2}
\left(\mn{B}+\widetilde{B}(t')\right)\left\{\int_0^{t'}\left(\mn{B}\widetilde{V}(t'')
- \mn{V}\widetilde{B}(t'')\right)\,\dd t''\right\}\,\dd t'.$$
We can in principle continue in the same way to find $q_k,\, k>2$.

Using the practical approximation for $B(t)$ and $V(t)$, we find $q_0 =
V_0/B_0$, $c_1 = 0$ and so 
$$q_1(t) = -\frac{V_0(B_0-B_1)}{B_0^2}\sin\omega t,\qquad
q_2(t) = \frac{V_0(B_0-B_1)}{4B_0^3}\left(B_1\cos 2\omega t - 4B_0\cos\omega t
+ 4B_0 - B_1\right) + c_2.$$
In what follows, we also need an expression for $\dot{q}_3$, which obeys
$\dot{q}_3 = -\omega B(t) q_2(t)/B_0$. Hence, we require a value for $c_2$,
which we again deduce by imposing the condition that
$$q_3 = -\frac{\omega}{B_0}\int_0^t \left(B_0 - B_1\cos\omega t'\right) \left\{K_1
\left(B_1 \cos 2\omega t' - 4B_0\cos\omega t' + 4B_0 - B_1\right) + c_2\right\}\,\dd t' + c_3,$$
where $K_1 = V_0(B_0-B_1)/4B_0^3$,
is bounded as $t\rightarrow\infty$. Since $\mn{\cos\omega t} = \mn{\cos\omega t \cos 2\omega t} = 0$
but $\mn{\cos^2\omega t} = 1/2$, this condition gives
$$c_2 = -\frac{V_0(B_0 - B_1)(4B_0 + B_1)}{4B_0^3}$$
and so
$$q_3(t) = \frac{V_0(B_0-B_1)}{24B_0^4}\left(3(8B_0^2 - 3B_1^2)\sin\omega t
- 9B_0B_1\sin 2\omega t + B_1^2\sin 3\omega t\right) + c_3.$$
To find the mean power, we again need to compute $\mn{\dot{q}}^2 =
\delta^2\mn{\dot{q}_1^2} + \delta^3 \mn{\dot{q}_1\dot{q}_2} + \delta^4\mn{2\dot{q}_1\dot{q}_3
+ \dot{q}_2^2} + O(\delta^6)$, where, as before, we observe that $\dot{q}_1(t)\dot{q}_2(t)$ is
an odd function of $t$ in the practical approximation, so its mean is zero.
From the above expressions for $q_1,\,q_2,\,q_3$, we find
$$\mn{\dot{q}_1^2} = \frac{1}{2}\left(\frac{V_0(B_0-B_1)}{B_0^2}\right)^2\omega^2,
\qquad
\mn{2\dot{q}_1\dot{q}_3 + \dot{q}_2^2} = -\frac{V_0^2(B_0-B_1)^3(B_0+B_1)}{2B_0^6}\omega^2.$$
Defining the mean power for large $R$, $\mn{p}_l$, as $\mn{p}_l := R\mn{\dot{q}^2}$ we have
\begin{eqnarray}
\mn{p}_l &=& \frac{1}{2}\left(\frac{V_0(B_0-B_1)}{B_0}\right)^2 R^{-1} -
\frac{V_0^2(B_0-B_1)^3(B_0+B_1)}{2B_0^2} \omega^{-2} R^{-3}
+ O\left(\omega^{-4} R^{-5}\right)\nonumber \\
&=& 42231.4 R^{-1} -3.67414\times10^{24}\omega^{-2} R^{-3} + O\left(\omega^{-4} R^{-5}\right)
\label{plarge}
\end{eqnarray}
where we have used the parameter values in Table~\ref{numvals} to obtain the last
expression.

\subsection{Comparison with numerics}
\label{cwn}

\begin{figure}[htbp]
\centering 
\includegraphics*[width=5.0in]{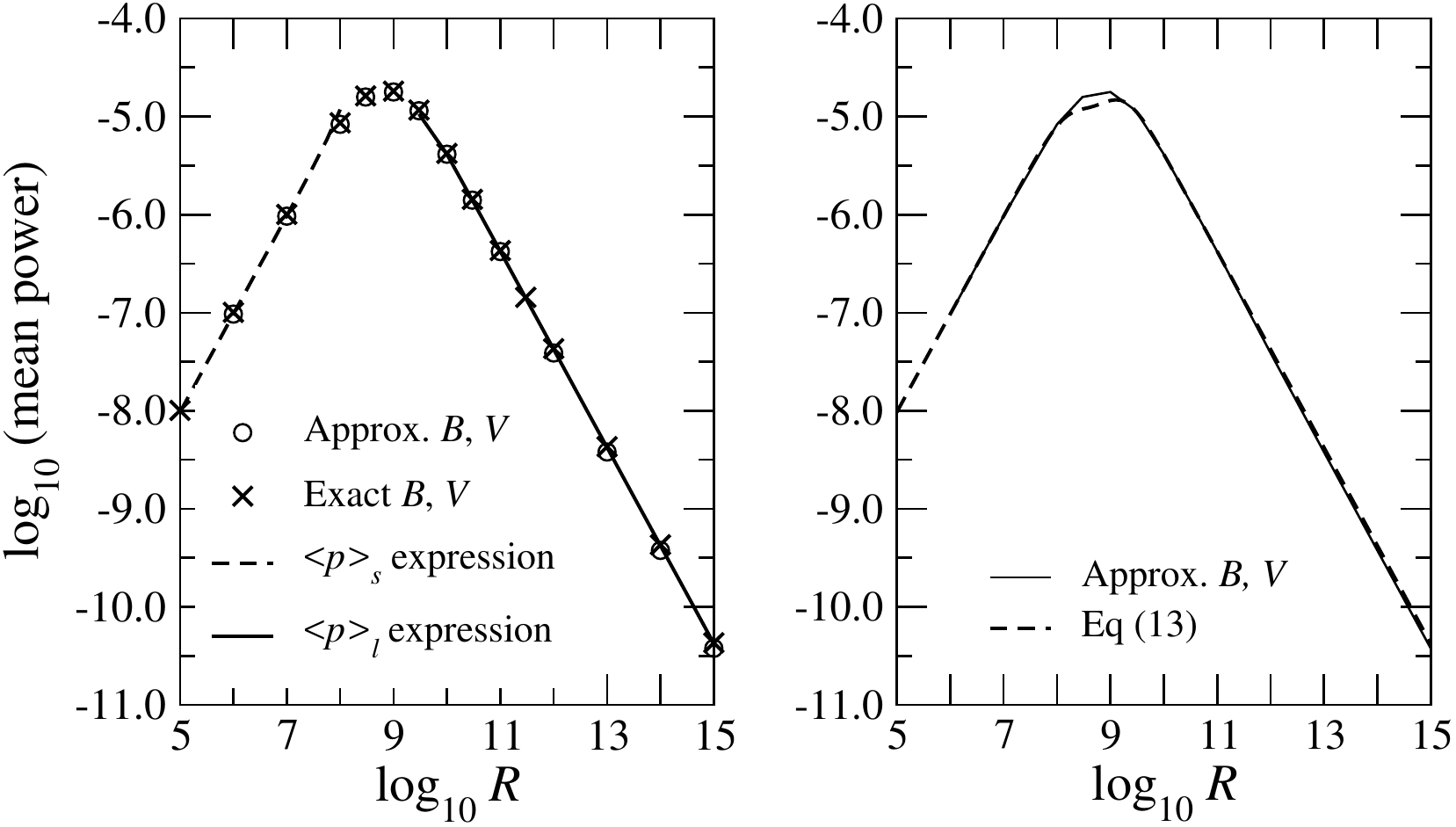}
\caption{Left: The mean power, computed directly from a numerically-obtained $i(t) = \dot{q}(t)$
waveform and not from Fourier series, versus $R$, using the approximate
(circles) and exact (crosses) expressions for $B(t)$ and $V(t)$. Also shown,
over appropriately restricted ranges of $R$, are the series/asymptotic expansions
for $\meps$ and $\mepl$, from equations~\eqref{psmall} and~\eqref{plarge} respectively.
Right --- see section~\ref{mpar}: mean power from the numerical $i(t)$ as in the left-hand figure with
approximate $B, V$, (solid line), and also using the rational approximation $\mn{p(R)}$,
equation~\eqref{ratapp} (dashed line).}
\label{pwr}
\end{figure} 

We now make two comparisons. The first, in Figure~\ref{pwr},
is a plot of the mean power, computed in four different ways:
\begin{itemize}
\item By solving the o.d.e.~\eqref{ode} numerically, using a good numerical algorithm
(the Livermore Stiff o.d.e.\ solver~\cite{lsode}), with the value of $q(0)$
computed by Fourier series,
using (a) the approximate $B(t)$, $V(t)$ functions (circles in Figure~\ref{pwr}) and
(b) the exact expressions for them (crosses);
\item By perturbation theory, as derived in the previous two subsections, with the dashed
line showing $\mn{p}_s$ for $R\in[10^5, 10^8]$, as given by equation~\eqref{psmall},
and the continuous line, $\mn{p}_l$ from equation~\eqref{plarge}, for
$R\in[3\times10^9, 10^{15}]$.
\end{itemize}
The agreement between approximate and exact
$B, V$ is seen to be very good, as are the perturbation expressions (within
their ranges of applicability).

\begin{figure}[htbp]
\centering 
\includegraphics*[width=5.0in]{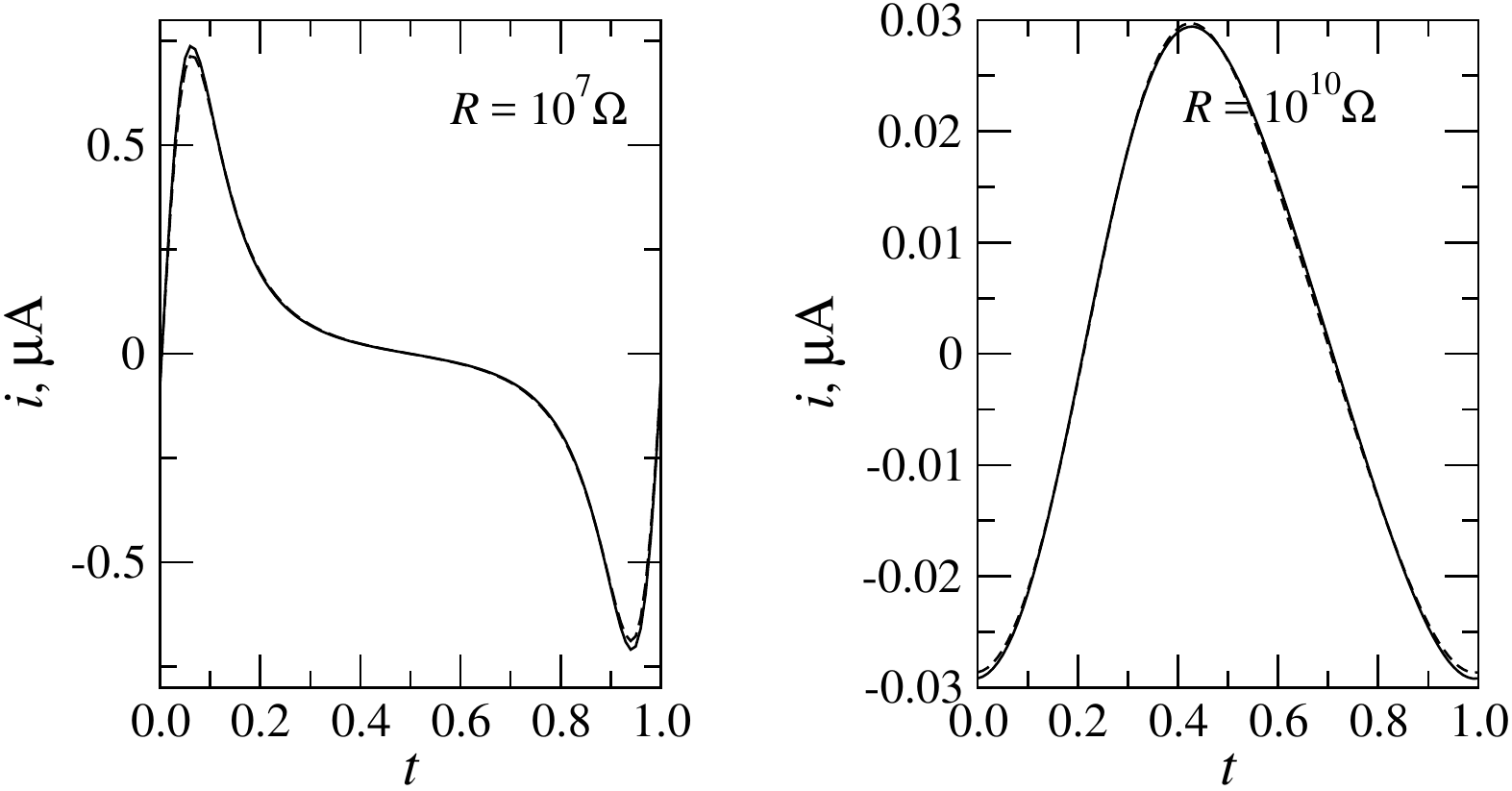}
\caption{The current, computed by solving the o.d.e.~\eqref{ode} numerically
(continuous lines) and using the perturbation series (dashed lines). On the left, $R =
10^7\,\Omega$ and the small $R$ series is used; on the right, $R = 10^{10}\,\Omega$ 
and the large $R$ series is used. The curves are sufficiently close that
they are almost indistinguishable.}
\label{icomp}
\end{figure} 

The second comparison is between currents, computed numerically using
the exact $B(t)$, $V(t)$, and from perturbation theory up to and including
the terms $\dot{q}_3(t)$, both in the small and large $R$ cases
($\dot{q}_3$ is only needed for large $R$). See
Figure~\ref{icomp}, which compares the numerical solution with the small
$R$ perturbation solution ($R = 10^7\,\Omega$) and with the large $R$
perturbation solution ($R = 10^{10}\,\Omega$). The two curves are almost
exactly superimposed.

\subsection{Applications of the perturbation solutions}
\label{appert}

We now briefly discuss some applications of equations~\eqref{psmall} and~\eqref{plarge},
the mean power expressions for small and large $R$ respectively.

In general, these expressions between them give a good approximation to $\mep$,
in a very simple form, for all $R$ except $10^8\leq R \leq 3\times 10^9$,
and this may in itself be useful. Furthermore, we have derived expressions
for $q(t)$ up to $O(\e^2)$ for small $R$ and $O(\delta^3)$ for large $R$,
where $\e = \omega R/\mn{B}$ and $\delta = 1/\e$. These approximations are
good, as can be seen from Figure~\ref{icomp}.

Less obviously, the perturbation series can also be used to estimate $\rpk$, the value of $R$
for which the peak mean power is obtained. Taking logarithms, and only including the first terms of
equations~\eqref{psmall} and~\eqref{plarge} respectively, we find
$$\ln\mn{p} = \ln\left(a_1\omega^2 R\right) + O\left(R^2\right)\qquad\mbox{and}\qquad
\ln\mn{p} = \ln\left(b_1 R^{-1}\right) + O\left(R^{-2}\right),$$
where $a_1$ and $b_1$ are given in~\eqref{a1xp} and~\eqref{plarge}
respectively. On a plot of $\ln\mn{p}$ versus $\ln R$, these are straight
lines, and their intersection point gives an estimate of $\rpk$.
This is
\begin{equation}
\rpk\approx \frac{1}{\omega}\sqrt{\frac{b_1}{a_1}} = 
\frac{B_0}{\omega}\left(1 - \frac{B_1^2}{B_0^2}\right)^{5/4}
= \frac{4.152}{\omega}\;\mbox{G}\Omega.
\label{intersect_rpk}
\end{equation}
We postpone discussion of the accuracy of this until we have 
a more accurate computation of $\rpk$ at the end of section~\ref{fccomp}.
Note that this argument will always overestimate the power, however.

\subsection{Mean power for all $R$?}
\label{mpar}

We end this section with an observation. Clearly, it would be useful to have
a simple formula for the mean power, $\mn{p(R)}$, valid for \textit{all} $R$, rather
than the restricted ranges accessible via perturbation theory.
A heuristic approach yields a non-rigorous result, which we now briefly discuss. 

Looking at equations~\eqref{psmall} and~\eqref{plarge} suggests that these
expressions might be, respectively, the Maclaurin series and an asymptotic
expansion of a single odd function of $\omega R$
$$\omega\, F(\omega R) = \omega\,\frac{A_1\cdot \omega R + A_3\cdot \omega^3 R^3}
{1 + B_2\cdot\omega^2 R^2 + B_4\cdot\omega^4 R^4},$$
where $A_1$, $A_3$, $B_2$ and $B_4$ are constants to be found. By
definition, this function is a candidate for $\mn{p(R)}$. In fact, all $F$
does is to interpolate smoothly between the small and large $R$ regimes;
how it behaves for intermediate values of $R$ is beyond our control.

Expanding $\omega F(\omega R)$ in a Maclaurin series, we find
$$\mn{p}_s = A_1\omega^2 R + (A_3 -A_1B_2)\omega^4 R^3 + O(\omega^6 R^5)$$
whereas the asymptotic expansion, that is, the Maclaurin series for
$\omega F\left(\frac{1}{\omega R}\right)$ in powers of $\frac{1}{\omega R}$, is
$$\mn{p}_l = \frac{A_3}{B_4}\frac{1}{R}  + \frac{A_1 B_4 - A_3 B_2}{B_4^2}
\frac{1}{\omega^2 R^3} + O(\omega^{-4} R^{-5}).$$
Matching the coefficients of the various powers of $R$ with
equations~\eqref{psmall} and~\eqref{plarge} gives four equations to solve
for $A_1$, $A_3$, $B_2$ and $B_4$, and doing so gives
\begin{equation}
\omega\,F(\omega R) 
= \omega^2 R\frac{2.45018\times 10^{-15} + 2.99655\times 10^{-34}\omega^2 R^2}
{1 + 6.75329\times 10^{-19}\omega^2 R^2 + 7.09553\times 10^{-39}\omega^4 R^4}
\stackrel{?}{=}\mn{p(R)}.
\label{ratapp}
\end{equation}
Figure~\ref{pwr}, right-hand side, shows a comparison between $\mn{p(R)}$ approximated in this
way, and a numerical computation. Solving $\partial\mn{p(R)}/\partial R = 0$ for $R$ 
gives one positive, one negative and four complex values of $R$. The
positive one is $R_{max} = 8.09\times 10^9/\omega = 1.29\times 10^9\Omega$ for $\omega
= 2\pi$, and $\mn{p(R_{max})} = 15.2\mu$W. Compare these with the
numerical computation in Figure~\ref{pwr}, which gives $R_{max} = 6.8\times
10^8\Omega$ and $\mn{p(R_{max})} = 18.2\mu$W.

\section{A Fourier series approach}
\label{FS}

An alternative approach to computing $q(t)$ is to use Fourier series:
having established in section~\ref{exist} that there is a unique attracting periodic
solution, the idea here is to expand it as a Fourier series. In this
section, we use the practical approximation for $B(t)$ and $V(t)$, although
the ideas could be applied for any $B$, $V$ of the same period. We start
by setting
\begin{equation}
q(t) = \sum_{k\in\mathbb Z}\alpha_k e^{\ii k\omega t},
\label{fser}
\end{equation}
where $\alpha_k\in\mathbb C$, with $\alpha_{-k} = \alpha_k^*$, so
$\alpha_0\in\mathbb R$, as is $q(t)$ for real $t$.
Substituting this in the o.d.e., we find that the $\alpha_k$ must obey the following relations:
\begin{subequations}
\label{recrel}
\begin{align}
\label{rra}
\alpha_1 & = \eta\alpha_0 -\alpha_{-1}- 2Q_0,\\ 
\label{rrb}
\alpha_2 & = \eta(1+\ii\e)\alpha_1 - \alpha_0 + Q_0,\\
\label{rrc}
\alpha_{k+1} & = \eta(1 + \ii\e k)\alpha_k - \alpha_{k-1},\qquad |k| \geq 2,
\end{align}
\end{subequations}
where we have used the dimensionless real parameters $\e = \omega R/B_0$,
as before, and $\eta := 2B_0/B_1 > 0$. Also, $Q_0 := V_0/B_1$, and both
$Q_0$ and $\alpha_k$ have dimensions of charge. In what follows, we 
sometimes need to distinguish between the two `exceptional equations',~\eqref{rra}
and~\eqref{rrb}, and the `general equation',~\eqref{rrc}, which is true for $|k| \geq 2$.
Here, there are just two exceptional equations because $B(t)$ and $V(t)$
are truncated at the first harmonic: the number of exceptional equations
grows linearly with the number of harmonics retained in $B(t)$ and $V(t)$.

There are two approaches to solving equations~\eqref{recrel}, and we explain these in the following two
subsections. In the first, we show an arithmetical solution
and in the second, we show how the same solution can be constructed using Bessel
functions, and we briefly discuss the asymptotic behaviour of this solution.

\subsection{Arithmetical solution}
\label{fccomp}

Here is a practical method for solving the difference equations~\eqref{recrel}
for $\alpha_k$, $k\in\mathbb Z$ and with $\alpha_{-k} = \alpha_k^*$.
There should be no free parameters in the solution, even though
solving~\eqref{recrel} is equivalent to solving the original o.d.e., whose
general solution contains one arbitrary constant. However, here
we seek only the periodic solution, $q(t)$, to the o.d.e., and not the
general solution, $q_c(t)$: only the latter contains an arbitrary constant.

The general solution to~\eqref{rrc}, which is of second order, 
has two arbitrary parameters. One just sets the scale, since, if
$\alpha_k$ is a solution, then so is $\lambda\alpha_k$ for any $\lambda$.
In fact, this scaling is set by either one of~\eqref{rra},~\eqref{rrb}. However, we still need
one more relation in order to pin down the second arbitrary parameter. By
analogy with the usual procedure for solving Mathieu's equation~\cite{a&s}, this
is furnished by considering the ratio $\rho_k := \alpha_k/\alpha_{k-1}$. 
 For $k \geq 2$, we have
$$\rho_{k+1} = \eta(1+\ii\e k) - \frac{1}{\rho_k} = d_k  - \frac{1}{\rho_k},$$
where $d_k := \eta(1+\ii\e k)$. Hence, $\rho_k = 1/(d_k - \rho_{k+1})$,
giving the continued fraction expansion
\begin{equation}
\rho_n = \frac{\alpha_n}{\alpha_{n-1}} = \frac{1}{d_n - \frac{1}{d_{n+1} - \frac{1}{d_{n+2}-\ldots}}}
\label{cf}
\end{equation}
From equation~\eqref{cf}, we can now compute $\rho_3$, which
depends on $\eta$, $\e$ but not on $\alpha_k$. Then we use equations~\eqref{rra}
and~\eqref{rrb}, and equation~\eqref{rrc} with $k=2$, along with the definition
of $\rho_3$, to find $\alpha_0,\ldots, \alpha_3$. That is, we solve
\begin{equation}
2\re\alpha_1 = \eta\alpha_0 - 2 Q_0,\;\;\;\alpha_2 = d_1\alpha_1 - \alpha_0 + Q_0,\;\;\;
\alpha_3 = d_2\alpha_2 - \alpha_1\;\;\;\mbox{and}\;\;\;\alpha_3 = \rho_3\alpha_2
\label{exeqs}
\end{equation}
for $\alpha_0,\ldots, \alpha_3$, obtaining
\begin{equation}
\alpha_0 = \frac{2Q_0(x+1)}{\eta + 2x},\;\;\;\alpha_1 = w(Q_0 - \alpha_0);
\label{exsol}
\end{equation}
then we use the second and third expressions in equation~\eqref{exeqs} to find $\alpha_2,\; \alpha_3$
in terms of $\alpha_0,\;\alpha_1$. Here, $w = (d_2 - \rho_3)/(1-d_1(d_2-\rho_3))$ and $x = \re w$.

Knowing $\alpha_2$, $\alpha_3$, we can now compute $\alpha_k$ for $k\geq 4$
from the general equation --- but not in the obvious way,
that is, by finding $\alpha_4$, followed by $\alpha_5$ and so on, since
iteration in this direction is unstable: the values of $\alpha_k$
so obtained rapidly become inaccurate, even for moderate values of $k$, and 
especially for large $R$. Instead, following section 19.28 in~\cite{a&s}, where an analogous
problem is solved, we iterate in the reverse direction: we fix a maximum value of $k$, $K$, 
say, and invert the general 
equation~\eqref{recrel} to give $\alpha_{k-1} = d_k\alpha_k - \alpha_{k+1}$.
We then use this to find, successively $\alpha_{K-j}$, $j = 1, \ldots, K-2$,
each in terms of $\alpha_K,\,\alpha_{K+1}$, which are, for the moment, unknown. At stage
$j$, we will have an expression of the form $\alpha_{K-j} = u_j \alpha_K + v_j \alpha_{K+1}$,
where $u_j,\, v_j\in\mathbb C$. Now, for $j = K-3$ and $K-2$, we have
$\alpha_3 = u_{K-3}\alpha_K + v_{K-3}\alpha_{K+1}$ and
$\alpha_2 = u_{K-2}\alpha_K + v_{K-2}\alpha_{K+1}$, and, since $\alpha_2$
and $\alpha_3$ are known, we can solve these two equations for $\alpha_K$
and $\alpha_{K+1}$.  With these now known, we can find $\alpha_k$, $k = 4, \ldots, K+1$,
and hence $q(t)$ (from equation~\eqref{fser}). In particular, $q(0)
\approx\sum_{|k|\leq K+1}\alpha_k$, which we used as our initial condition in
the numerical computation of $q(t)$ in section~\ref{cwn}.

\begin{figure}[htbp]
\centering 
\includegraphics*[width=4.8in]{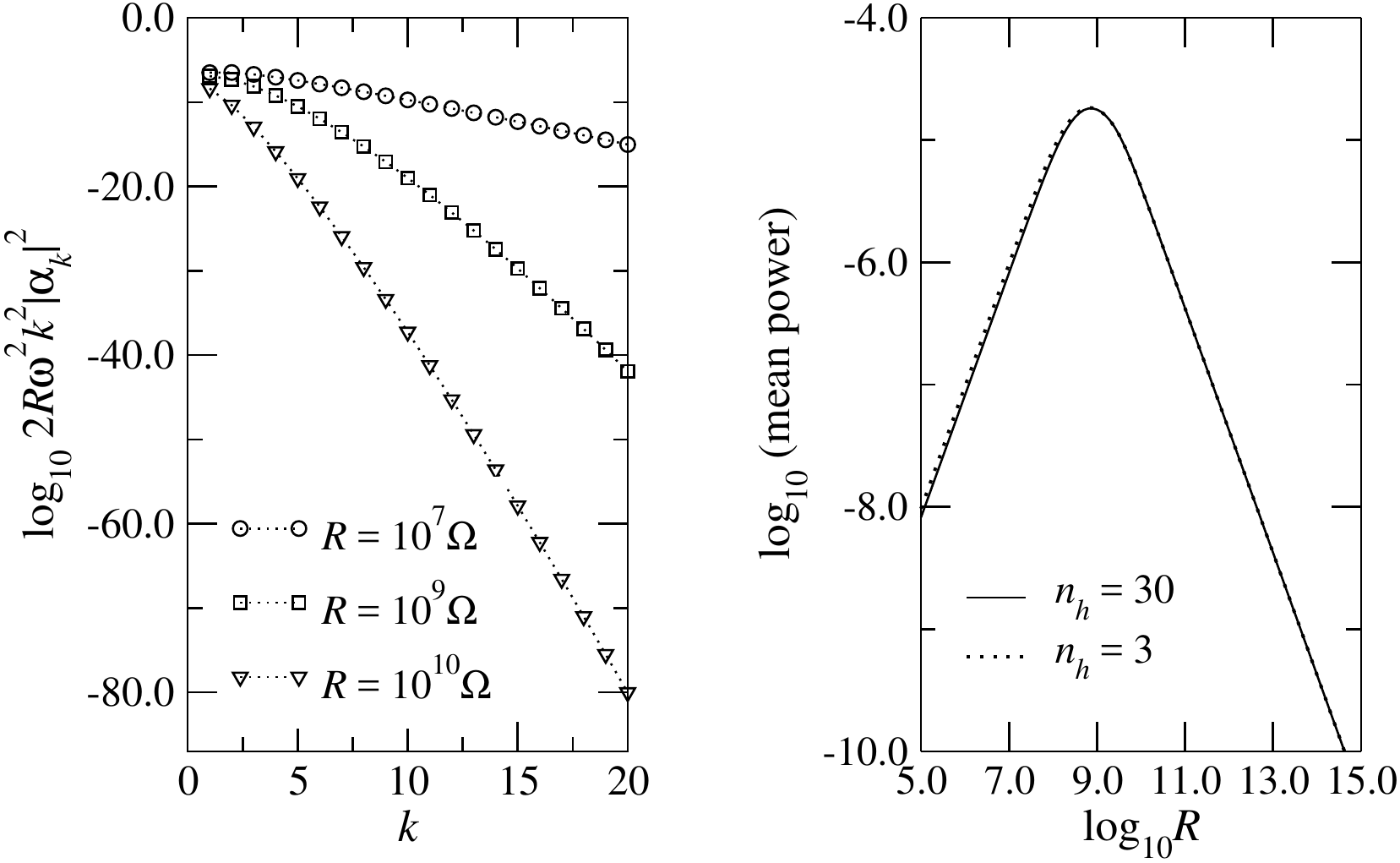}
\caption{Left: A logarithmic plot of the energy in the $k$-th harmonic, $\log_{10}
2R\omega^2 k^2|\alpha_k|^2$, $k = 1,\ldots, 20$, computed as described 
in section~\ref{fccomp}, for $R = 10^7, 10^9$ and $10^{10}\Omega$.
Right: Mean power as a function of $R$ computed from $2\omega^2 R\sum_{k=1}^{n_h} k^2|\alpha_k|^2$,
with $n_h = 30$ and $n_h = 3$. Compare with Figure~\ref{pwr}.}
\label{fcof_pwr}
\end{figure} 

Figure~\ref{fcof_pwr} shows a logarithmic plot of the modulus of the first 21 Fourier
coefficients so obtained, in the cases $R = 10^7, 10^9$ and $10^{10}\Omega$.
For comparison, we compute the mean power $\mn{p(R)}$ by
(a) $2\omega^2 R\sum_k k^2|\alpha_k|^2$; (b) by solving the
o.d.e.~\eqref{ode} numerically, as was done for Figure~\ref{pwr}; and (c) by
using the approximate formulae~\eqref{psmall},~\eqref{plarge}, if they
apply. The agreement is very good --- see Table~\ref{pwr_comp}.
\begin{table}[h!]
\centering
\begin{tabular}{|l|c|c|c|}\hline
\multicolumn{1}{|c}{\raisebox{2.5ex}{ }Method} & \multicolumn{1}{|c}{$R=10^7\Omega$} &
\multicolumn{1}{|c|}{$R=10^9\Omega$} & \multicolumn{1}{|c|}{$R=10^{10}\Omega$}\\ \hline
(a) Sum of Fourier coefficients  & $0.9652 \mu$W  & $17.80 \mu$W & $4.132 \mu$W\\
(b) Numerical solution of o.d.e  & $0.9652 \mu$W  & $17.80 \mu$W & $4.132 \mu$W\\
(c) Eq.~\eqref{psmall} or~\eqref{plarge} if valid & $0.9652 \mu$W & --- & $4.130 \mu$W\\\hline
\end{tabular}
\caption{Comparison of the mean power computed in three different ways, for
three different values of $R$. The first value in row (c) comes from
equation~\eqref{psmall}, the third from~\eqref{plarge}. Neither formula
applies for $R = 10^9\Omega$.}
\label{pwr_comp}
\end{table}

We now make a practical point. In general, the values of
$|\alpha_k|$ decrease very rapidly with $k$, the more so with increasing
$R$. In fact, to a good approximation, for $R\in [10^5, 10^{15}]\Omega$, we can use 
\begin{equation}
\mn{p(R, n_h)} := 2\omega^2 R\sum_{k = 1}^{n_h} k^2|\alpha_k|^2,
\label{p_from_FS}
\end{equation}
for a small number of harmonics, $n_h$. For an accurate estimate of the mean power, we use $n_h = 30$.
However, even with $n_h = 3$, using equations~\eqref{exeqs} and~\eqref{exsol} to find $\alpha_1$,
$\alpha_2$ and $\alpha_3$, the largest relative error between $\mn{p(R, 3)}$ and
$\mn{p(R, 30)}$ is about 15\%; this largest relative error occurs for $R\in[10^5,
3\times 10^7]$ and is approximately constant over this range --- see
Figure~\ref{fcof_pwr}. For other values of $n_h$, we find the following maximum relative errors:
$n_h = 4:\;6\%$, $n_h = 5:\;2.5\%$, $n_h = 7:\;0.3\%$.

\begin{figure}[htbp]
\centering 
\includegraphics*[width=4.8in]{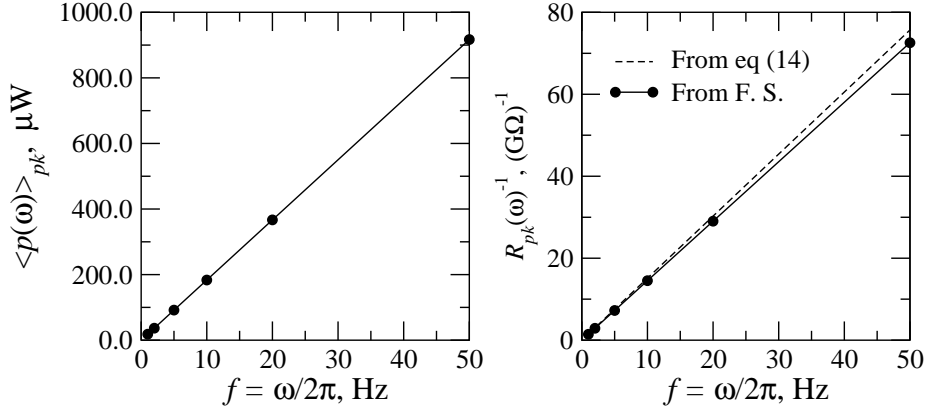}
\caption{Fourier-series-based power computations. Left: the peak mean power, $\mepkw$, $\mu$W,
as a function of frequency $f = \omega/2\pi = 1, 2, 5, 10, 20$ and 50 Hz.
Right: $\rpk(\omega)^{-1}$, the value of load resistance at which this
peak mean power is obtained, plotted for the same values of $\omega$ (solid
line), alongside the estimate of $\rpk$ given in equation~\eqref{intersect_rpk}
(dashed line). The range of $\rpk$ itself, as opposed to its reciprocal, is 
about $14$--$690$ M$\Omega$.}
\label{maxpwr}
\end{figure}

With an algorithm to compute the Fourier coefficients now in place, we can
do more however. For instance, we can study the effect of
$\omega$ on the power output, and we include in Figure~\ref{maxpwr} a plot of the peak mean power
obtained as $R$ varies, as a function of $\omega$. Specifically, we compute
the mean power, $\mn{p(R, n_h)}$, from~\eqref{p_from_FS} with $n_h = 30$, which is large
enough to ensure that the error is completely negligible, and then vary $R$ to find
the maximum value of the mean power, $\mepkw$. That is, $\mepkw := \max_{R> 0}\mn{p(R, 30)}$.
We then denote by $\rpk(\omega)$ the value of $R$ for which $\mepkw$ is obtained.
A very strong linear trend is noticeable in both $\mepkw$ and $\rpk(\omega)^{-1}$,
and using the data in Figure~\ref{maxpwr}, we find $\mepkw\approx
2.92\omega\;\mu$W and $\rpk(\omega)\approx 4.33/\omega$~G$\Omega$.
The latter should be compared with equation~\eqref{intersect_rpk}, which
predicts that $\rpk(\omega)\approx 4.15/\omega$~G$\Omega$ for the parameter
values in Table~\ref{numvals} --- a very good agreement.

The fact that $\rpk(\omega)$ is proportional to $1/\omega$ suggests the
following way of understanding qualitatively the behaviour of the circuit in Figure~\ref{Matt}. If we 
replace $C(t)$ by a constant capacitance $\ceff$ and compute the power
transferred to load $R$, we find that it is a maximum when $R = \rpk =
1/(\omega\ceff)$. Now, from above, we have that $\rpk(\omega)\approx 4.33\times 10^9/\omega$,
suggesting that $\ceff\approx 2.31\times 10^{-10}$~F. Looking now at Figure~\ref{CV},
left, we see that $\ceff$ lies between the maximum and minimum values of
$C(t)$. It is similar, but not equal to the mean value of $C(t)$, which is
about $1.05\times 10^{-10}$~F, and it is clear that the actual value of
$\ceff$ can only be computed from an accurate power computation, such as
that carried out here using Fourier series.

\subsection{A solution based on Bessel functions}

Along with equations~\eqref{recrel}, consider also
\begin{equation}
\beta_{k+1} + \beta_{k-1} = \eta(1+\ii\e k)\beta_k, \qquad k\in\mathbb Z
\label{beteq}
\end{equation}
and
\begin{equation}
\gamma_{k+1} + \gamma_{k-1} = \eta(1-\ii\e k)\gamma_k, \qquad k\in\mathbb Z.
\label{gameq}
\end{equation}
The Bessel function of the first kind~\cite{a&s}, $J_\nu(z)$, is defined for $\nu,\, z\in\mathbb C$,
with $\nu$ referred to as the order, and this function obeys
\begin{equation}
J_{\nu+1}(z) + J_{\nu-1}(z) = \frac{2\nu}{z}J_\nu(z).
\label{Ceq}
\end{equation}
In the light of this, solutions to equation~\eqref{beteq}
can be expressed in terms of a Bessel function of the appropriate (complex) order.
For this to work, we require that $\eta(1+\ii\e k) = 2\nu/z$ for all $k$, where $z$ is
independent of $k$, and $\nu = k + \ii\xi$ for $\xi\in\mathbb R$. Taking
these together, we see that $z\eta(1+\ii\e k) = 2k + 2\ii\xi$, and so $z$
must be purely imaginary. Hence $\nu = k -\ii/\e$ and $z = -2\ii/\eta\e$,
giving $\beta_k = C W_k$ where $W_k := J_{k-\ii/\e}(-2\ii/\eta\e)$ and $C$
is an arbitrary constant. Note
that $J_{\nu}(z)$ is bounded as $\hbox{Re}(\nu) \to +\infty$, as shown in
Appendix~II, Property~\ref{jbig}. Analogously, for any $k\in\mathbb Z$, equation~\eqref{gameq}
has the solution $\gamma_k=D Z_k$, where $D$ is an arbitrary constant and
$Z_k:=J_{k+\ii/\e}(2\ii/\eta \e)$.

Now set $\delta_k=\beta_k$ for $k\ge 1$ and $\delta_k=\gamma_{-k}$ for $k\le -1$.
Then, for $k\ge 2$, we have
$$ \delta_{k+1} + \delta_{k-1} = \beta_{k+1} + \beta_{k-1} = \eta \left( 1 + \ii\e k \right) \beta_{k} =
\eta \left(1 + \ii\e k \right) \delta_{k},$$
while, for $k\le -2$,
$$\delta_{k+1} + \delta_{k-1} = \gamma_{-(k+1)} + \gamma_{-(k-1)} = \gamma_{-k-1} + \gamma_{-k+1} 
= \eta \left(1 + \ii\e k \right) \gamma_{-k} = \eta \left( 1 + \ii \e k \right) \delta_{k},$$
so that, comparing the last two equations with~\eqref{rrc}, we see that, if we set
$\alpha_k = \delta_{k}$ for all $|k| \ge 1$, we obtain a solution to~\eqref{rrc} depending on the
complex constants $C$ and $D$.

Now, Property~\ref{jstar} in Appendix~II states that $J_{\nu^*}(z^*) = \left( J_{\nu}(z) \right)^*$.
Hence, setting $D=C^*$, for all $k\ge 1$, we have
\begin{equation}
\alpha_{k}^* =  \beta_k^* = C^* W_k^* = C^* \left(J_{k-\ii/\e}(-2\ii/\eta \e)\right)^* =
D J_{k+\ii/\e}(2\ii/\eta \e) = D Z_{k} = \gamma_{k} = \al_{-k}
\label{aka-k}
\end{equation}
so that $\al_{k}^*=\al_{-k}$ for all $k \neq 0$. So, if we require $q(t)$ to be real,
we are left with only one free parameter $C\in\mathbb C$.

To compute $C$, we go back to equations~\eqref{rra},~\eqref{rrb}. Solving
these for $\alpha_0$ gives
\begin{equation}
\alpha_0 = \frac{CW_1 + C^* W_1^* + 2Q_0}{\eta} = \eta(1+\ii\e)CW_1 - C W_2 + Q_0,
\label{al0eq}
\end{equation}
where we have used $\alpha_i = CW_i$ for $i = 1, 2$. The right-hand equality can be rewritten
in the form $z_1 C + z_2 C^* = r$, where $z_1, z_2, C\in\mathbb C$ and
$r\in\mathbb R$. Specifically, $z_1 = W_1\left(1 - \eta^2(1+\ii\e)\right)\!/\eta + W_2$,
$z_2 = W_1^*/\eta$ and $r = Q_0(1-2/\eta)$. By considering $z_1 C + z_2 C^* = r$ and its
complex conjugate, we can solve for $C$ to obtain $C = r(z_1^* - z_2)/\Delta$, where
$\Delta = |z_1|^2 - |z_2|^2$. Numerical evidence indicates that
$\Delta\in(0, 1)$ for all $R > 0$, so $C$ exists for all positive $R$.

The expression $\alpha_k = C J_{k-\ii/\e}(-2\ii/\eta\e)$, $k\geq 1$, with
$\alpha_0$ being given by~\eqref{al0eq}, is in principle good for any values of the
parameters. In practice, however, the method for computing $\alpha_k$ in 
Section~\ref{fccomp} is often useful, especially for small $R$ (leading to small
$\e = \omega R/B_0$, which requires the evaluation of $J_{k-\ii/\e}(z)$ for small $\e$
--- algorithms to do this tend to be slow, especially when $|k-\ii/\e|\approx |z|$).

One advantage of expressing the Fourier coefficients in terms of Bessel
functions is that we immediately see from our expression for
$\alpha_k$, along with~\eqref{mJas} in Appendix~II, that
$$|\alpha_k| \sim \left(\frac{e}{\eta\e}\right)^k \frac{1}{k^{k+\frac{1}{2}}},$$
so clearly the Fourier series for $q(t)$ converges for all non-zero $\e, \eta$.

\section{Conclusions and further work}

We have studied the periodically-excited triboelectric nanogenerator (TENG) from a mathematical
viewpoint. Our main aim has been to derive expressions for the mean power as a
function of load resistance $R$ and excitation frequency $\omega$, although
as a by-product, we also compute the current, from
which other quantities of interest can be derived.

The TENG has a single state variable, the charge, $q(t)$, and the time
evolution of $q(t)$ is described by the linear, first-order, non-autonomous o.d.e.\ in
equation~\eqref{ode}. This has a periodically-varying coefficient, the reciprocal
capacitance $B(t)$, which makes analysis of 
the problem less straightforward than, say, the piezoelectric device discussed in~\cite{zhang},
the o.d.e.\ for which, while still non-autonomous, has constant coefficients.

After proving that the o.d.e.~\eqref{ode} has a unique periodic attractor, we
derive perturbation series for $q(t)$ in the cases
of small and large $R$, equations~\eqref{psmall} and~\eqref{plarge}
respectively. We give general expressions for the first few
coefficients in these series before using them to estimate the current, $i(t) = \dot{q}(t)$,
and the mean power, $R\times$[mean squared current over one period].
Comparison with numerics shows these expressions to be good for all $R$
except $R\in[10^8, 3\times 10^9]\Omega$ (approximately), these values
of $R$ being neither `small' nor `large' in the context of this problem. 
However, by using a simple argument based on the intersection of straight lines,
the two perturbation series can be used to estimate $\rpk$, the value of
$R$ that maximises the mean power, and this approach results in a simple expression,
equation~\eqref{intersect_rpk}, for $\rpk(\omega)$.

Since the o.d.e.\ has a unique periodic solution $q(t)$, we then discuss its
Fourier expansion. We give two procedures to find the Fourier
coefficients, $\alpha_k$, for $k$ as large as desired. The first is an
algorithm to find $\alpha_k$, which requires only simple computing
machinery, and the second is an expression for $\alpha_k$ in terms of Bessel
functions of the first kind, $J_\nu(z)$. From $\alpha_k$, the mean
power, for any $R$ and $\omega$, can be computed to any degree of precision, as can $q(t)$ and $i(t)$
--- from the latter, the peak current and peak power can then be found if
desired. One important aspect of the connection with Bessel functions is that we
can deduce the behaviour of $\alpha_k$ for large $k$.

Our chief interest has been the mean power as a function of $R$ and
$\omega$ and we discuss how this may be estimated by a single expression,
valid for \textit{all} $R$. We propose two possibilities. The first is
heuristic and comes from an observation about the two
series, equations~\eqref{psmall} and~\eqref{plarge}: we show that if these
series are both expansions of the same rational function, then we can
compute an approximation to this function explicitly --- see equation~\eqref{ratapp}.
This estimate is simple to use but entirely heuristic, although it does compare 
favourably with the exact result.

The second possibility comes from our study of the Fourier coefficients in
terms of Bessel functions, which shows that they decrease in magnitude
rapidly (as $1/k!$ in fact) with increasing index $k$. In practice, even the first three
coefficients are enough to give a reasonable power estimate (relative error
$\leq 15\%$) for all $R$.

Clearly, the power output also depends on other parameters in the problem,
for example the thickness $x_1$, $x_2$ of the dielectrics, the excitation amplitude
$z_0$ and the triboelectric charge density $\sigma_T$. A study of the
effects of these parameters and others has recently been published in~\cite{aem}.

This work raises some interesting questions for further investigation. For
example, in the TENG context, the fundamental frequencies of $B(t)$ and $V(t)$ have to
be the same. What can we say about solutions to o.d.e.s like~\eqref{ode}
where this is not the case? What more can we say about the convergence of the
perturbation series? Can we say anything about the the behaviour of the
modulus of the Fourier coefficients in the case where $B(t)$ and/or $V(t)$
are not approximated as truncated Fourier series? What can we say in the
case of a load that is not purely resistive?

\bigskip
\bigskip

\centerline{\Large\bf Appendix I --- derivation of $I(x)$}

\bigskip
This appendix gives the mathematical part of the derivation of the function $I(x)$
that features in the o.d.e.\ studied in this paper, and it is included for the sake
of completeness. The context is a device known as a triboelectric generator, consisting
of a pair of parallel rectangular plates whose sides are of length $l$ and $w$.

The definition of $I(x)$ is the following indefinite integral:
$$I(x) := \int\arctan\left(\frac{l/w}{2(x/w)\sqrt{4 (x/w)^2 + (l/w)^2 +1}}\right)\,\dd x.$$
It arises as a consequence of Gauss' law, and we have taken the integral directly
from~\cite{envsci17}, which should be consulted for further details.
Setting $\alpha
= l/w$ and substituting $y = x/w$, this becomes
\begin{equation}
I(y) = w\int\arctan\left(\frac{\alpha}{2y\sqrt{4y^2 + \alpha^2 + 1}}\right)\,\dd y.
\label{orig_int}
\end{equation}
Perhaps surprisingly, this integral can be expressed in closed form for all
$\alpha > 0$, provided that $x > 0$. The first step to showing this is to multiply the
integrand by 1 and integrate by parts, giving
$$\frac{I}{w} = y\arctan\left(\frac{\alpha}{2y\sqrt{4y^2 + \alpha^2 + 1}}\right)
- \int y \frac{d}{dy}\arctan\left(\frac{\alpha}{2y\sqrt{4y^2 + \alpha^2 +
1}}\right)\,\dd y.$$
We then use
$$\frac{d}{dy}\arctan\left(\frac{1}{g(y)}\right) = \frac{-g'(y)}{1 + g^2(y)},$$
and with $g(y) = (2y/\alpha)\sqrt{4y^2 + \alpha^2+1}$, we find
$$\frac{I}{w} = y\arctan\left(\frac{\alpha}{2y\sqrt{4y^2 + \alpha^2 + 1}}\right)
+2\alpha \int\frac{y(8y^2 + \alpha^2+1)}{\sqrt{4y^2 +
\alpha^2+1}\left[4y^2(4y^2+\alpha^2+1) + \alpha^2\right]}\,\dd y.$$
Now define
$$I_1 = \int\frac{y(8y^2 + \alpha^2+1)}{\sqrt{4y^2 + \alpha^2+1}\left[4y^2(4y^2+\alpha^2+1) + \alpha^2\right]}\,\dd y$$
and substitute $u^2 = 4y^2+\alpha^2+1$. This gives
$$I_1 = \int\frac{y(2u^2 - \alpha^2 - 1)}{u(4u^2y^2 + \alpha^2)}\frac{u}{4y}\,\dd u
= \frac{1}{4}\int\frac{2u^2-\alpha^2-1}{4u^2y^2 + \alpha^2}\,\dd u,$$
and using $4y^2 = u^2 - \alpha^2 - 1$, we find
$$I_1 = \frac{1}{4}\int\frac{2u^2 - \alpha^2 - 1}{u^2(u^2 - \alpha^2-1)+\alpha^2}\,\dd u
= \frac{1}{4}\int\frac{2u^2 - \alpha^2 - 1}{(u^2 - \alpha^2)(u^2-1)}\,\dd u $$
$$ = \frac{1}{4}\int\frac{(u^2 - \alpha^2) + (u^2-1)}{(u^2 - \alpha^2)(u^2-1)}\,\dd u 
= \frac{1}{4}\int\frac{1}{u^2-1} + \frac{1}{u^2 - \alpha^2}\,\dd u.$$
Expressing both these terms in partial fractions, we then obtain
$$I_1 = \frac{1}{8}\ln\frac{u-1}{u+1} + \frac{1}{8\alpha}\ln\frac{u-\alpha}{u+\alpha}.$$
Hence, finally, we have
\begin{equation}
I(x) = x\arctan\frac{\alpha w}{2xu} + \frac{w}{4}\left(\alpha\ln\frac{u-1}{u+1} + \ln\frac{u-\alpha}{u+\alpha}\right),
\label{Idef}
\end{equation}
where $u = \sqrt{4(x/w)^2 + \alpha^2+1}$.

Points to note:
\begin{itemize}
\item The expressions for $I(x)$ and $I_0$ are unchanged if $l$
and $w$ are swapped (which also replaces $\alpha = l/w$ with $\alpha^{-1}$). This
is, of course, exactly as it should be: it cannot matter which way round we
label the sides of the rectangular plates.
\item Since both $\arctan x > 0$ and $\alpha/(2x\sqrt{4x^2+\alpha^2+1}) > 0$
for $x>0$, $I(x)$ is the integral of a strictly positive function, and so
$I(x)$ is a monotonically increasing function of $x > 0$.
Hence, $G(t)$ as defined in equation~\eqref{ode_orig} is
always positive provided only that $x_1 + x2 > 0$; and so $B(t) = G(t)/A > 0$
for all $t$.
\end{itemize}

We shall also need $I_0 = \lim_{x\rightarrow 0} I(x)$, which is
$$I_0 = \frac{\alpha w}{4}\ln\left(\frac{\alpha^2+2-2\sqrt{\alpha^2+1}}{\alpha^2}\right)
+\frac{w}{4}\ln\left(2\alpha^2+1-2\alpha\sqrt{\alpha^2+1}\right).$$
When $\alpha = 1$, this gives $I_0 = (w/2)\ln(3 - 2\sqrt 2) \approx -0.881374 w$.

\bigskip
\bigskip

\centerline{\Large\bf Appendix II --- properties of $\pmb{J_\nu(z)}$}

\bigskip
\begin{property}
Fix $u\neq 0\in\mathbb R$ and $z\in\mathbb C$. Then, as $k\rightarrow\infty$,
$\left|J_{k+\ii u}(z)\right| \rightarrow 0.$
\label{jbig}
\end{property}
\begin{proof}
Equations~\eqref{Jas} and~\eqref{stir} below are taken from~\cite{a&s}, \cite{tranter}.
Equations~\eqref{Jas} and~\eqref{stir}, and all the asymptotic expansions in this proof,
have an error term which multiplies the right-hand sides by a factor $(1 + O(|\nu|^{-1}))$.
We start from the asymptotic expansion
\begin{equation}
J_\nu(z) \sim \frac{1}{\Gamma(\nu+1)} \left({\frac{z}{2}}\right)^\nu.
\label{Jas}
\end{equation}
The latter is valid for fixed $z\in\mathbb C$ and
complex $\nu$ with $|\nu|\gg |z|$. Stirling's approximation for large $|\nu|$ is
\begin{equation}
\Gamma(\nu + 1) \sim \sqrt{2\pi\nu}\left(\frac{\nu}{e}\right)^\nu.
\label{stir}
\end{equation}
We put $\nu = k + \ii u$, where $k\in\mathbb Z$, and set
$\theta = \arg\nu$. With these definitions, applying Stirling's approximation
to equation~\eqref{Jas}, we obtain
$$J_\nu(z) \sim \frac{1}{\sqrt{2\pi\nu}}\left(\frac{ze}{2\nu}\right)^\nu
= \frac{1}{\sqrt{2\pi}}\left(\frac{ze}{2}\right)^{\ii u}\times
e^{u\theta}\, e^{-\ii u\ln|\nu|} e^{-\ii\theta(k+\frac{1}{2})}
\left(\frac{ze}{2}\right)^k\frac{1}{|\nu|^{k + \frac{1}{2}}}.$$
Only terms to the right of the $\times$ symbol depend on $k$. Furthermore, we can replace
$e^{u\theta}$ with unity, since $u$ is fixed, so
$\theta = \lim_{k\rightarrow\infty} \arg (k+\ii u) = 0$.
Hence, taking the modulus,
\begin{equation}
\left|J_{k+\ii u}(z)\right| \sim C \left(\frac{|z|e}{2}\right)^k\frac{1}{k^{k+\frac{1}{2}}},
\label{mJas}
\end{equation}
where $C>0$ is independent of $k$. The property is then established by letting $k\rightarrow\infty$.
\end{proof}

\bigskip
\begin{property}
For $\nu, z\in\mathbb C$, $J_{\nu^*}(z^*) = \left(J_v(z)\right)^*$.
\label{jstar}
\end{property}
\begin{proof}
This follows from the series representation~\cite{a&s} of the Bessel function
$$J_{\nu}(z) = \sum_{m=0}^{\infty} \frac{(-1)^m}{m!\,\Gamma(m+\nu+1)}
\left(\frac{z}{2}\right)^{2m+\nu}$$
along with the property that $\Gamma(z^*)=(\Gamma(z))^*$ of the Gamma function~\cite{a&s}
and the fact that $(z^*)^{\al^*} = (z^\al)^*$. The latter can be seen by
writing $z = re^{\ii\theta}$, $\alpha = a + \ii b$ and expanding.
\end{proof}

\end{document}